\documentclass[12pt,aaspp4]{aastex}

\begin{document}

\title{A Survey for ``Normal'' Irregular Satellites Around Neptune: Limits to Completeness\altaffilmark{1}}  
\author{Scott S. Sheppard\altaffilmark{2}}    

\affil{Carnegie Institution of Washington, Department of Terrestrial Magnetism, \\
5241 Broad Branch Rd. NW, Washington, DC 20015 \\ sheppard@dtm.ciw.edu}

\and

\author{David Jewitt and Jan Kleyna}
\affil{Institute for Astronomy, University of Hawaii, \\
2680 Woodlawn Drive, Honolulu, HI 96822 \\ jewitt@ifa.hawaii.edu, kleyna@ifa.hawaii.edu}

\altaffiltext{1}{Based largely on data collected at Subaru Telescope, which is operated by the National Astronomical Observatory of Japan.}
\altaffiltext{2}{The observations for this work were acquired while the author was at the Institute for Astronomy at the University of Hawaii.}

\begin{abstract}  

We surveyed 1.75 square degrees of sky near Neptune to an R-band $50
\%$ detection efficiency of 25.8 mags (corresponding to radii of about
17 km for an assumed albedo of 0.04).  We discovered one new outer
satellite, Psamathe (S/2003 N1), about 20 km in radius and having a
distant retrograde orbit and moderate eccentricity.  Until 2003
Neptune was only known to have two satellites which exhibited orbital
signatures indicative of capture.  Both of these, Triton and Nereid,
are unusual when compared to the irregular satellites of other giant
planets.  With recent discoveries of four additional satellites by
Holman et al. (2004) it is now apparent that Neptune has a distant
``normal'' irregular satellite system in which the satellites have
radii and orbital properties similar to those of the satellites of
other giant planets.  We find that the satellite size distribution at
Neptune is not well determined given the few objects known to date,
being especially sensitive to the inclusion of Triton and Nereid in
the sample.  Finally, we note that Psamathe and S/2002 N4 have similar
semi-major axes, inclinations and eccentricities.  They may be
fragments of a once larger satellite.

\end{abstract}


\section{Introduction}

The irregular satellites of the giant planets have moderate to high
orbital eccentricities and inclinations with distant prograde or
retrograde orbits.  Because of their extreme orbits they are believed
to have been captured (Kuiper 1956).  This is unlike the regular
satellites which are close to their respective planets with circular,
low inclination, prograde orbits that probably formed within
circumplanetary disks of gas and dust as part of the planet formation
process.

Energy must be dissipated for initially unbound satellites to become
captured by a planet (Everhart 1973).  With no energy dissipation, a
temporarily captured satellite will either be ejected from the system
or impact the planet within a few centuries (the best recent example
of this was provided by D/Shoemaker-Levy 9).  Three possible capture
mechanisms have been discussed, but none operate efficiently in the
modern Solar System: 1) drag due to gas around the forming planet
(Pollack, Burns \& Tauber 1979; McKinnon \& Leith 1995) 2) pull-down
capture as the planet's mass grows (Heppenheimer \& Porco 1977) and 3)
collisional or gravitational interactions between asteroids and/or
satellites moving within the planet's Hill sphere (Colombo \& Franklin
1971; Tsui 2000; Astakhov et al. 2003; Agnor \& Hamilton 2004).

The irregular satellite systems of Jupiter, Saturn and Uranus show
remarkable similarities in their populations, size distributions and
orbital properties (Sheppard \& Jewitt 2003; Sheppard, Jewitt and
Kleyna 2005; Jewitt \& Sheppard 2005).  These similarities are hard to
understand in view of the vast differences between the formation of
the gas and ice giant planets.  Gas giants Jupiter and Saturn most
likely formed by core accretion (Pollack et al. 1996) or disk
instabilities (Boss 2001).  Ice giants Uranus and Neptune have about
ten times less mass, are deficient in H and He compared to the gas
giants, and must have formed differently from the gas giants
(e.g. Lissauer 1995; Thommes et al. 2002; Boss 2002).  Gas drag is
unlikely to have been effective at Uranus and Neptune because these
planets have little gas.  Pull down capture likewise is unlikely
because the ice giants had no runaway growth in mass caused by
hydrodynamic inflow of nebular H and He.  Instead, the preferred
capture mechanism is through collisional or gravitational interactions
between small bodies within the Hill spheres of the planets.  Such
three-body interactions are independent of the planet formation
scenario and mass and could have operated around both gas and ice
giants (Jewitt \& Sheppard 2005).

Neptune's satellite system is unusual compared to those of the other
giant planets because it has no massive regular satellites.  The
current regular satellites of Neptune are less than 5 Neptune radii
away from the planet and the largest (Proteus) is only about 200 km in
radius, almost an order of magnitude smaller than the largest regular
satellites of the other giant planets.  A possible reason is that the
very massive retrograde satellite Triton, probably a captured Kuiper
Belt object, ejected any regular satellites which were beyond about 5
Neptune radii (Goldreich et al. 1989).  In fact, even the regular
satellites currently observed may have been disrupted by Triton in the
past and what we observe today is the reaccumulation of those
fractured bodies (Banfield \& Murray 1992).  In addition, Triton may
have scattered Nereid into its highly eccentric orbit.  Because of
Nereid's small semi-major axis and low inclination compared to other
irregular satellites it is suspected to have once been a regular
satellite of Neptune (Goldreich et al.1989).

\begin{figure}
\epsscale{0.7}
\plotone{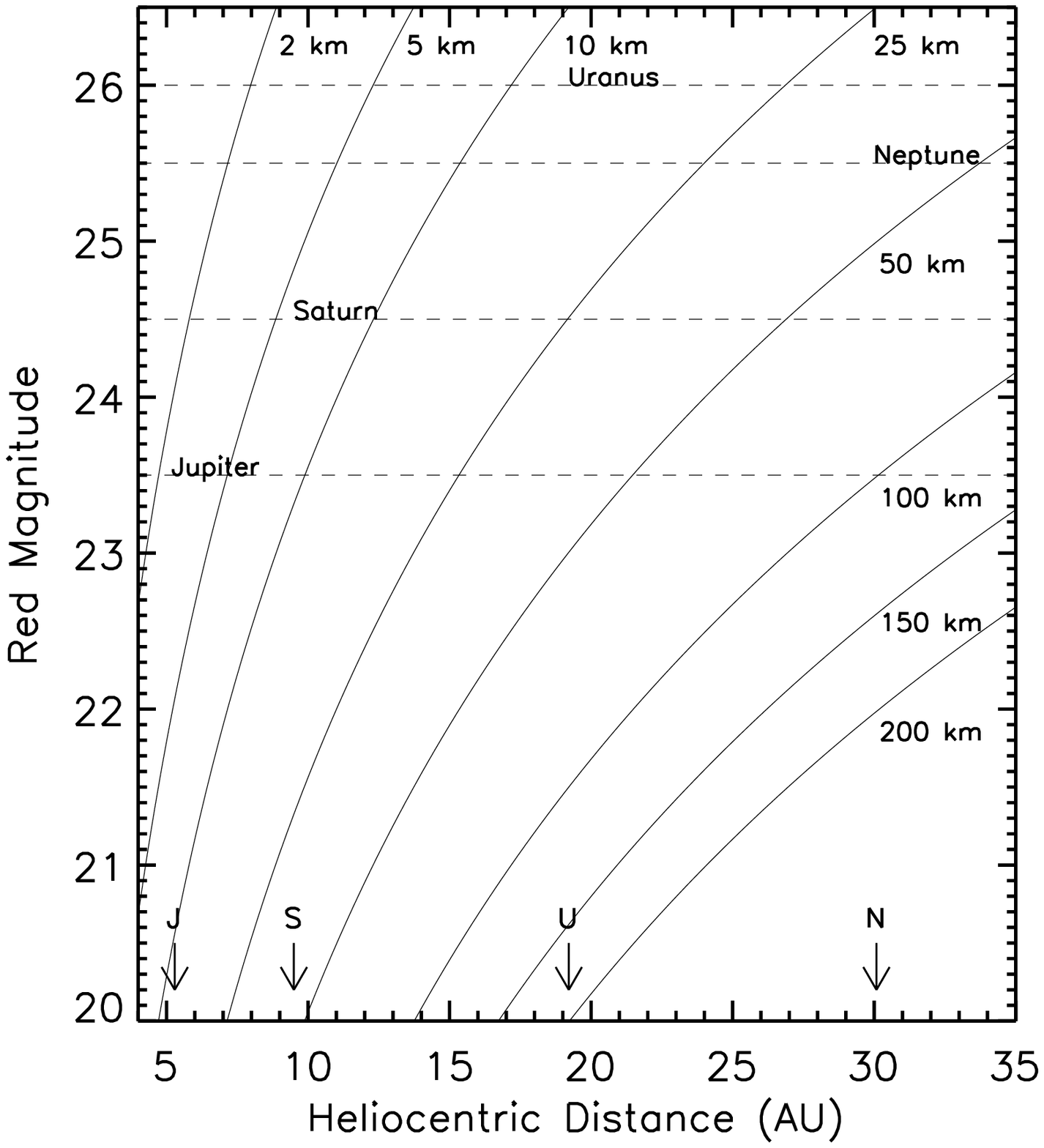}
\caption{The distances of the planets versus the observable small body
population diameter (solid lines) for a given red magnitude assuming
an albedo of 0.04.  Dashed lines show the approximate survey magnitude
completeness limits for satellites of each planet to date.  Though
Jupiter satellite surveys are the shallowest of the four planets they
have been the most sensitive to small satellites because of Jupiter's
closer proximity to Earth.}
\label{fig:distance26neptune} 
\end{figure}

Because of Neptune's extreme distance (Figure
\ref{fig:distance26neptune}) it has the least well-characterized outer
irregular satellite system.  We wish to determine if the ice giant
Neptune has a population of small, outer, irregular satellites similar
to those of gas giants Jupiter and Saturn and fellow ice giant Uranus.
Until recently Neptune was not known to have any of what we will call
``normal'' outer irregular satellites.  Only the ``unusual'' Nereid
was known and has a relatively close in, very eccentric, low
inclination orbit.  Nereid also happens to be the largest known outer
satellite of any planet.  Holman et al. (2004) recently surveyed $1.4$
square degrees around Neptune to a $50 \%$ detection efficiency of
$m_{R} \sim 25.5$ and announced four small, outer irregular satellites
of Neptune; S/2002 N1-N4 (Holman et al. 2003a; 2003b).  Here we
discuss an independent survey to slightly fainter magnitudes and
covering a slightly larger area.

\section{Observations and Analysis}

We surveyed the space around Neptune when it was near opposition.  The
geometry of Neptune in the survey is shown in Table~1.  The
observations were obtained near new moon on UT August 29 and 30, 2003
using the Suprime-Cam camera on the Subaru 8.2 meter diameter
telescope atop Mauna Kea.  The Suprime-Cam imager uses 10 MIT/LL $2048
\times 4096$ CCDs arranged in a $5 \times 2$ pattern (Miyazaki et
al. 2002) and with $15 \micron$ pixels that give a scale of $0.\arcsec
20$ pixel$^{-1}$.  The field-of-view is about $34\arcmin \times
27\arcmin$ with the North-South direction along the long axis.  Gaps
between the chips are about $16 \arcsec$ in the North-South direction
and only $3 \arcsec$ in the East-West direction.

The images were obtained through a Kron-Cousins R-band filter with the
telescope autoguided sidereally.  Image reduction was performed by
first bias subtracting and then flat-fielding with twilight flats.
Seeing during the two nights varied from $0.\arcsec 45$ to $0. \arcsec
7$ FWHM.  Objects at Neptune's distance trailed about $0.\arcsec 45$
during the 500 second exposures.  Landolt (1992) standards were used
for calibration on both photometric nights.

The region where planetary satellites may be stable is known as the
Hill sphere where the radius, $r_{H}$, depends on the planet's mass
and distance from the Sun as

\begin{equation}
r_H = a_p \left[\frac{m_p}{3M_{\odot}}\right]^{1/3}
\label{eq:hill}
\end{equation}

where $a_p$ and $m_p$ are the orbital semi-major axis and mass of the
planet and $M_{\odot}$ is the mass of the sun.  Table 2 shows the Hill
radii for the outer planets.

\begin{figure}
\epsscale{0.6}
\plotone{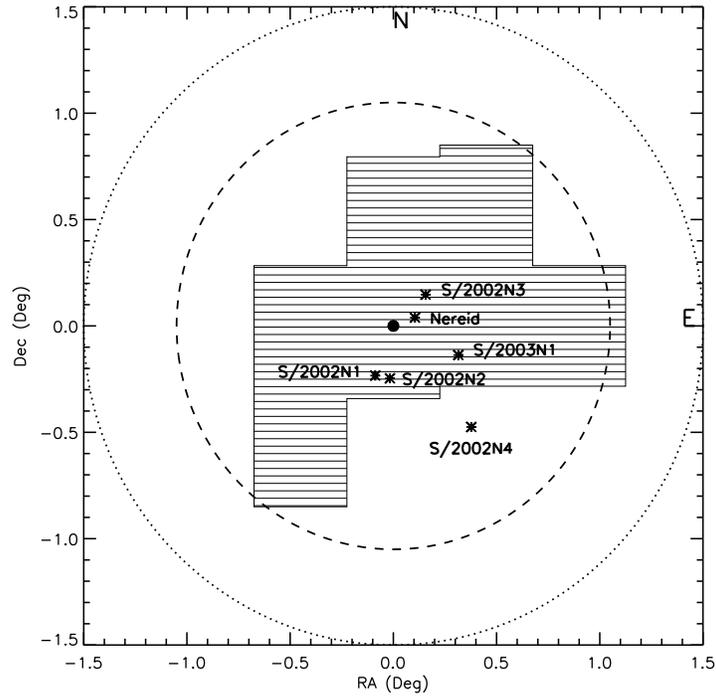}
\caption{The survey area around Neptune searched for satellites using
the Suprime-Cam on the 8.2m Subaru telescope.  The black dot at the
center represents Neptune's position.  Stars represent the positions
at the time of observations of the outer satellites of Neptune.  The
dotted circle shows the projected Hill sphere of Neptune while the
dashed circle shows the theoretical outer limits of stability for
Neptune satellites (at 0.7 $r_{H}$).}
\label{fig:areaneptune} 
\end{figure}

The area of the Hill sphere searched for satellites is shown in
Figure~\ref{fig:areaneptune}.  Seven fields were imaged 3 times each
on one night and 2 times each on the second night for a total of 5
images per field or 35 images for the survey.  The second night's
fields were at the same angular distance from Neptune as those from
the first night but the background star fields were slightly different
because of Neptune's movement between the two nights.  Images of each
field were spaced by about 33 minutes on a given night.  Approximately
1.75 square degrees around Neptune were observed, not accounting for
chip gaps and bright stars.  The image of Neptune was positioned in a
gap between the CCD chips to prevent saturation of the detectors.

\begin{figure}
\epsscale{0.5}
\plotone{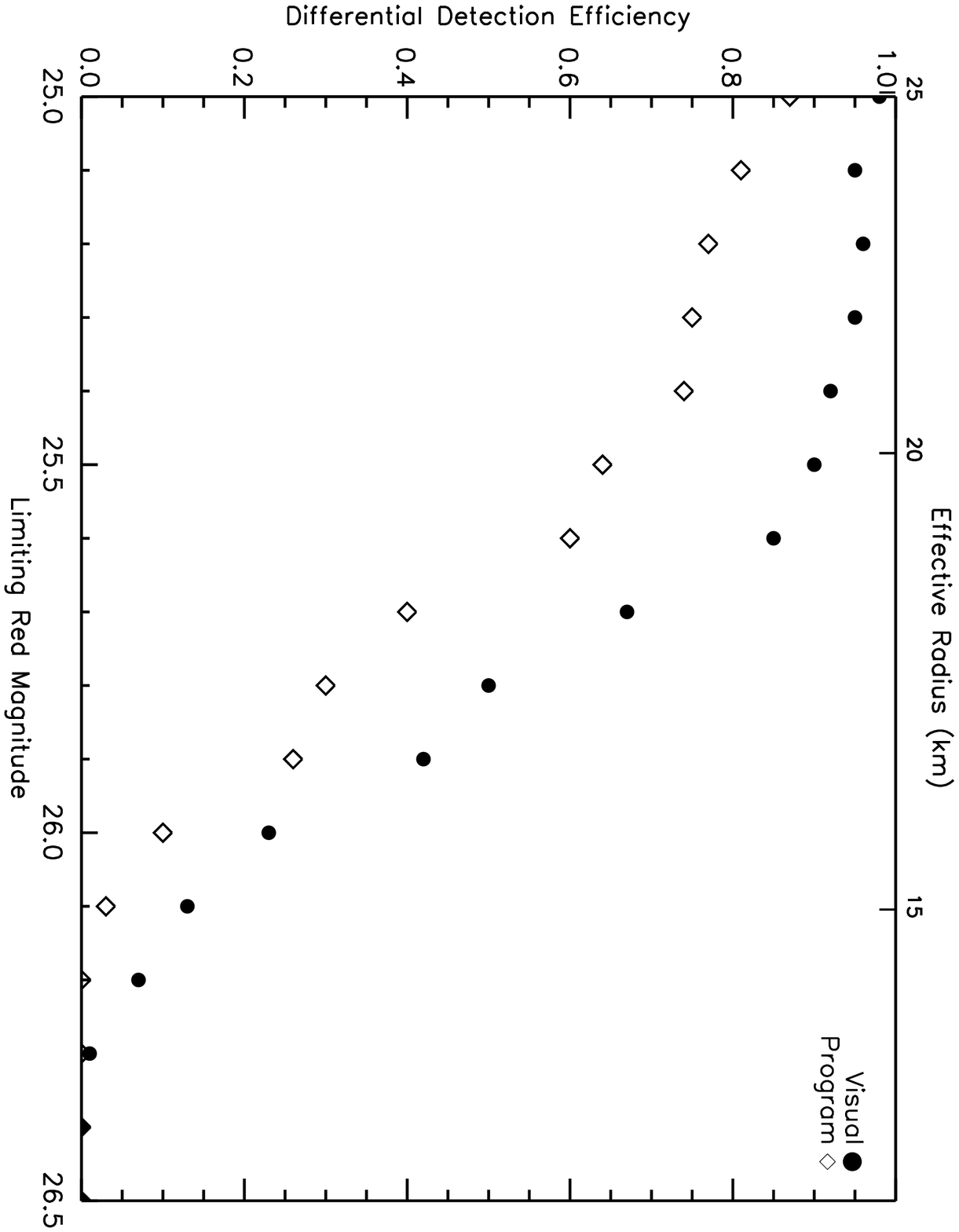}
\caption{Detection efficiency of the Neptune survey versus the
apparent red magnitude.  The $50 \%$ detection efficiency is at about
25.8 mags as determined from visual blinking and 25.7 mags determined
from a computer program.  All fields were searched with both
techniques.  The efficiency was determined by placing artificial
objects matched to the Point Spread Function (PSF) of the images with
motions similar to Neptune in the survey fields.  Effective radii of
the apparent magnitude was calculated assuming the object would have
an albedo of 0.04.  The efficiency does not account for objects which
would have been undetected because of the chip gaps.  Scattered light
was not a significant problem in the survey.}
\label{fig:effneptune} 
\end{figure}

We searched for Neptune satellites in two complementary ways.  A
computer algorithm was used to detect objects which appeared in all
three images from one night and had a motion consistent with being
beyond the orbit of Jupiter (motion of 18 to $1$ arcsecond per hour).
Second, all fields were searched a second time by displaying them on a
computer screen and visually blinking them for any slow moving
objects.  The limiting magnitude of the survey was determined by
placing artificial objects in the fields matched to the point spread
function of the images with motions mimicking that of Neptune ($\sim
3.5$ arcseconds per hour).  Results are shown in Figure
\ref{fig:effneptune} for both the visual blinking and computer
algorithm.  The visual blinking was slightly more efficient with a $50
\%$ detection efficiency at an R-band limiting magnitude of about 25.8
magnitudes, which we take as the limiting magnitude of this survey.

There was virtually no scattered light beyond about 45 arcseconds from
Neptune.  Scattered light did not significantly affect our detection
efficiency until about 20 arcseconds from Neptune at which point the
background was only $30 \%$ higher than the nominal sky background.
The region within $\sim$10 arcseconds of Neptune fell in a chip gap
and was unobservable.

\section{Results and Discussion}

Through this survey we discovered one new Neptune satellite, Psamathe
(S/2003 N1), which was reported on the IAU Circular Number 8193
(Sheppard et al. 2003).  Holman et al. (2004) detected Psamathe on
only one night in their 2001 survey but did not originally obtain a
second night and thus were unable to confirm this object as a
satellite of Neptune.  We also recovered, without prior knowledge of
their locations, S/2002 N1, S/2002 N2 and S/2002 N3 as well as Nereid.
The only other known outer satellite of Neptune, S/2002 N4, was not in
our fields.  All five new outer satellites of Neptune now have well
determined orbits as a result of observations of each taken over
several years by Holman et al. (2004) and our group.

We relate the apparent red magnitude of an object, $m_{R}$, to its
radius, $r$, through

\begin{equation}
r = \left[ \frac{2.25\times 10^{16}R^{2}\Delta ^{2}}{p_{R}\phi
(\alpha)} \right]^{1/2} 10^{0.2(m_{\odot} - m_{R})}
\label{eq:appmagneptune}
\end{equation}

\noindent in which $r$ is in km, $R$ is the heliocentric distance in
AU, $\Delta$ is the geocentric distance in AU, $m_{\odot}$ is the
apparent red magnitude of the sun ($-27.1$), $p_{R}$ is the red
geometric albedo, and $\phi (\alpha)$ is the phase function in which
the phase angle $\alpha=0$ deg at opposition.  We assume $\phi
(\alpha) = 10^{-0.4 \beta \alpha}$, where $\beta$ is the ``linear''
phase coefficient.  Using Equation \ref{eq:appmagneptune}, data from
Table~1 and an albedo of 0.04 we find that our $50 \%$ detection limit
at 25.8 magnitudes corresponds to a satellite with radius of $17
(\frac{0.04}{p_{R}})^{1/2}$ km.

\begin{figure}
\epsscale{0.6}
\plotone{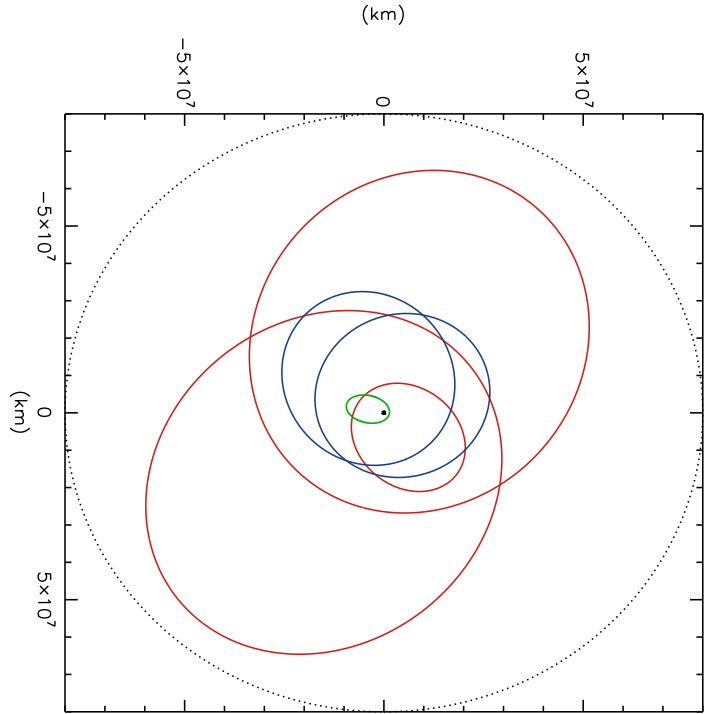}
\caption{A plan view of Neptune's satellite orbits.  The three small
outer retrograde irregular orbits are shown in red, the two small
outer prograde irregular orbits are shown in blue.  Nereid's orbit is
shown in green.  Triton's orbit is barely visible on this scale and is
represented by the black dot at the center.  The dashed circle shows
the theoretical outer limit of stability for Neptune satellites at 0.7
$r_{H}$.  The orbits are projected into the ecliptic, centered on
Neptune with axes in kilometers with zero degrees longitude on the
left side of the x-axis.}
\label{fig:planview} 
\end{figure}

The radius of Psamathe is about 20 km if we assume an albedo of 0.04.
Psamathe is in a retrograde orbit with an inclination of 137 degrees
with respect to the ecliptic and an eccentricity of 0.45. The
semi-major axis of Psamathe is about $46\times 10^{6}$ km which
corresponds to $0.4 r_{H}$.  The relatively large eccentricity allows
Psamathe to reach almost $0.6 r_{H}$ from Neptune
(Figure~\ref{fig:planview}), near the theoretical stable limit of $0.7
r_{H}$ for retrograde satellites (Hamilton \& Krivov 1997).

\begin{figure}
\epsscale{0.5}
\plotone{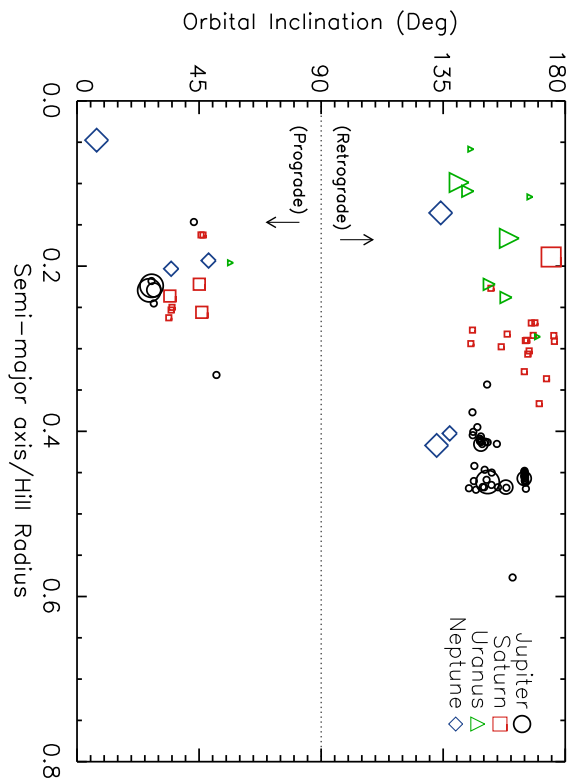}
\caption{An inclination comparison between the currently known 97
irregular satellites of the giant planets.  The horizontal axis is the
fraction of the satellite's mean semi-major axis compared to its
respective planet's Hill radius.  The vertical axis is the mean
inclination of the satellite to the ecliptic.  The size of the symbol
represents the radius of the object: Large symbol $r > 25$ km, medium
symbol $25 > r > 10$ km, and small symbol $r<10$ km.  All giant
planets independent of their mass or formation scenario appear to have
similar irregular satellite systems.  The ``unusual'' irregular Nereid
is seen in the lower left.  Triton has been omitted since its
inclination is only defined with respect to Neptune's equator since
tidal evolution has probably modified its inclination.  The new
irregular satellites discovered in the past few years around Neptune,
including Psamathe (S/2003 N1), are similar to the other known
irregular satellites of the giant planets.  All regular satellites
would fall near the origin of this plot.}
\label{fig:irrsats} 
\end{figure}

\begin{figure}
\epsscale{0.5}
\plotone{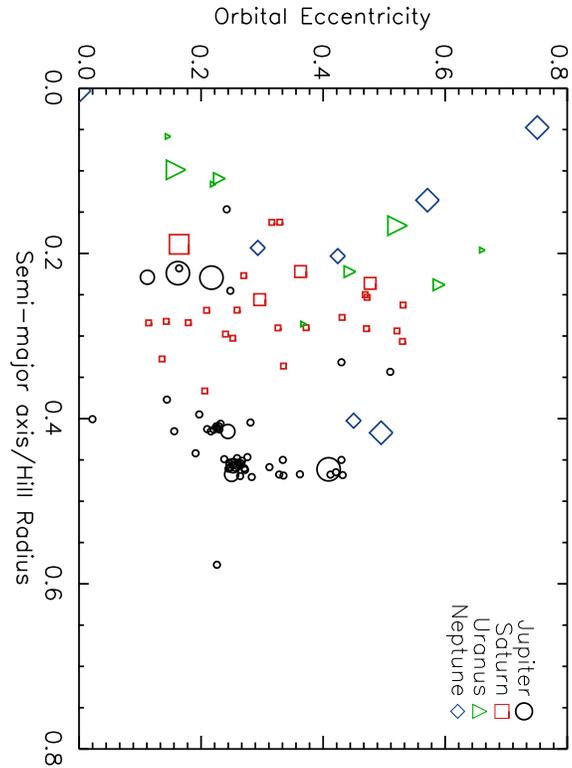}
\caption{Same as Figure \ref{fig:irrsats} except eccentricity is
plotted on the vertical axis.  Both Triton and Nereid standout in this
plot.  Nereid is in the upper left while Triton is located at the
origin of the plot.  The five newly discovered outer satellites of
Neptune are very similar to the known irregulars around the other
giant planets.  All regular satellites would fall near the origin of
this plot.}
\label{fig:irrsatsecc} 
\end{figure}

We list the properties of all the known outer satellites of Neptune in
Table 3.  Figures \ref{fig:irrsats} and \ref{fig:irrsatsecc} compare
the semi-major axes with inclinations and eccentricities,
respectively, of the irregular satellites of all the planets.  Both
Nereid and Triton standout in these figures leading us to label them
as ``unusual'' irregular satellites.  Nereid is quite large relative
to other irregular satellites and has the lowest inclination as well
as one of the smallest semi-major axes and largest eccentricities
compared to the rest of the known outer irregular satellites of the
giant planets.  Triton is almost an order of magnitude larger and has
over an order of magnitude smaller semi-major axis than other
irregular satellites and also has a circular orbit, likely
significantly modified by tidal interactions with Neptune (Goldreich
et al. 1989; Chyba et al. 1989).

The Neptune satellites discovered by Holman et al. (2004) and in this
work are ``normal'' irregular satellites as judged by their large
semi-major axes and orbital eccentricities and inclinations.  The
Neptune irregular satellites Psamathe (S/2003 N1) and S/2002 N4 have
similar large semi-major axes, inclinations and eccentricities and
thus may be daughter satellites of a once larger parent, as also
mentioned by Holman et al. (2004), but further refinement of the
orbits is needed before anything definitive can be said.  Further
discoveries may reveal more small satellites which share similar
semi-major axes and inclinations as part of a dynamical family like
those observed at Jupiter (Sheppard and Jewitt 2003).  Families have
also been reported at Saturn (Gladman et al. 2001) but these appear
significant only in inclination space, unlike the satellites of
Jupiter which are grouped in \textit{both} inclination and semi-major
axis.  No other obvious groupings are apparent.  The inclination
region $60 < i < 140$ degrees is void of known satellites consistent
with the action of the Kozai instability (Kozai 1962; Carruba et
al. 2002; Nesvorny et al. 2003).

\subsection{Size Distribution}

\begin{figure}
\epsscale{0.6}
\plotone{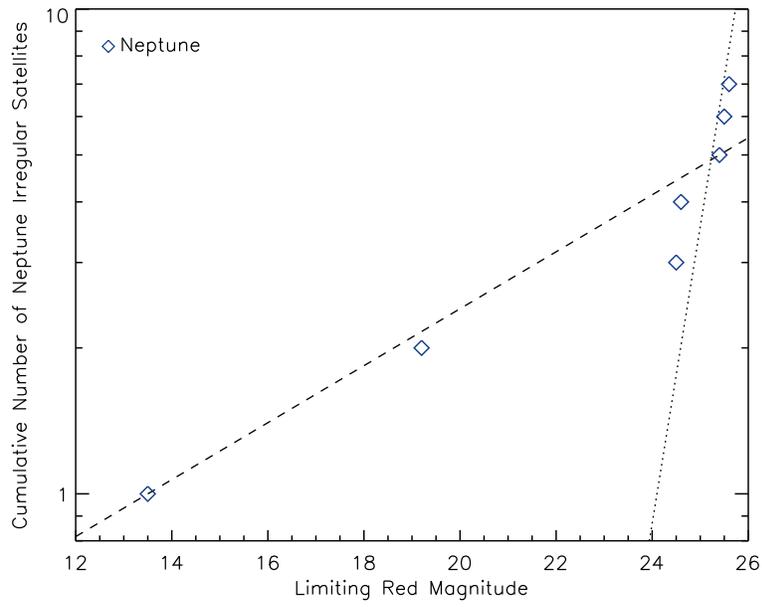}
\vspace{1cm}
\caption{The cumulative luminosity function (CLF) for satellites of
Neptune with orbits indicative of capture.  The dashed line shows the
best fit of the CLF using all seven satellites ($\alpha \sim 0.06$).
The dotted line shows the best fit using only the five small outer
irregular satellites with $r < 100$ km ($\alpha \sim 0.6$). Further
data are needed since the CLF is very sensitive to the few bright
objects.}
\label{fig:clfnep} 
\end{figure}

We represent the cumulative luminosity function (CLF), which describes
the sky-plane number density of objects brighter than a given
magnitude, by

\begin{equation}
\mbox{log}[\Sigma (m_{R})]=\alpha (m_{R}-m_{o}) \label{eq:slope}
\end{equation}

\noindent where $\Sigma (m_{R})$ is the number of objects per unit
area brighter than $m_{R}$, $m_{o}$ is the magnitude zero point, and
10$^{\alpha}$ describes the slope of the luminosity function.
Figure~\ref{fig:clfnep} shows the CLF using all seven known Neptune
satellites which have orbits indicative of capture.  We believe that
the outer satellites of Neptune are complete to near 25.5 mags ($r >
20$ km) through our survey, the Holman et al. (2004) survey and
additional null result surveys (Gladman et al. 2000).  See Table 2 for
the expected completeness limits for the outer irregular satellites of
the planets.

\begin{figure}
\epsscale{0.6}
\plotone{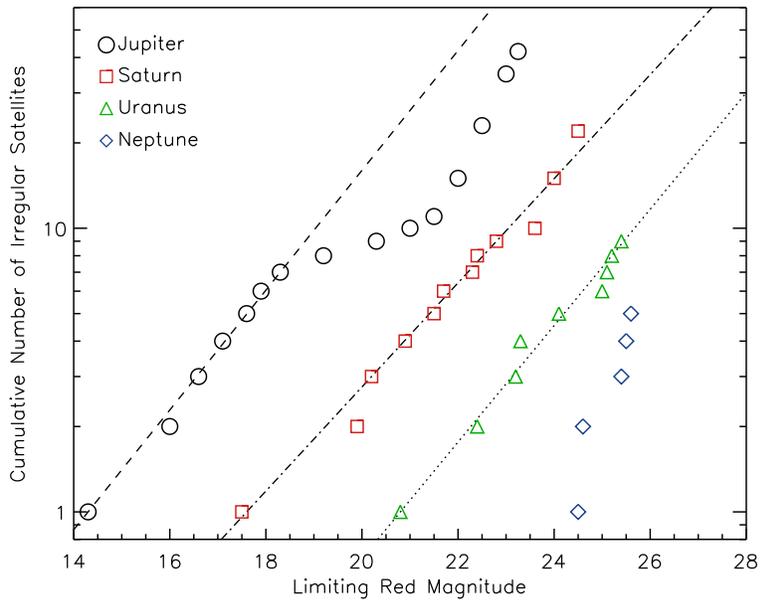}
\vspace{1cm}
\caption{The cumulative luminosity function (CLF) for the outer
irregular satellites with $r < 100$ km around Jupiter, Saturn, Uranus
and Neptune.  For clarity and in order to compare similar sized outer
irregular satellites we have omitted Neptune's Triton and Nereid which
are plotted in Figure~\ref{fig:clfnep}.  The slopes for irregular
satellites with $100 > r > 10$ km are plotted for Jupiter, Saturn and
Uranus.  They all are shallow and very similar ($\alpha \sim 0.20$)
but because of the different distances the further planets CLF's are
shifted to the right.  Neptune's irregulars with $r < 100$ km appear
to have a steeper slope but if Nereid ($r \sim 170$ km) and/or Triton
($r \sim 1350$ km) are added the slope becomes much shallower.  See
the text for details.}
\label{fig:cumnept} 
\end{figure}

Including all seven satellites around Neptune which have orbits
indicative of capture we find $\alpha \sim 0.06$, but this result is
not significant in the sense that it is extremely sensitive to the
inclusion of the ``unusual" irregulars Triton and Nereid.  We directly
compare similarly sized irregular satellites ($r<100$ km) of all the
giant planets in Figure~\ref{fig:cumnept}.  Neptune's irregular
satellites with $10 < r < 100$ km (which excludes Triton and Nereid)
show $\alpha = 0.6 \pm 0.1$ and $m_{o} = 24.5 \pm 0.4$ while including
Nereid gives $\alpha \sim 0.1$.  The sensitivity of the slope to the
inclusion or exclusion of Nereid shows that further discoveries are
needed in order to obtain a reliable CLF for Neptune's outer irregular
satellites.  To date, the results are broadly consistent with the
$\alpha \sim 0.2$ found for the irregular satellites with $10 < r <
100$ km around the other giant planets Jupiter, Saturn, and Uranus
(Sheppard \& Jewitt 2003; Kavelaars et al. 2004; Sheppard et
al. 2005).

\begin{figure}
\epsscale{0.6}
\plotone{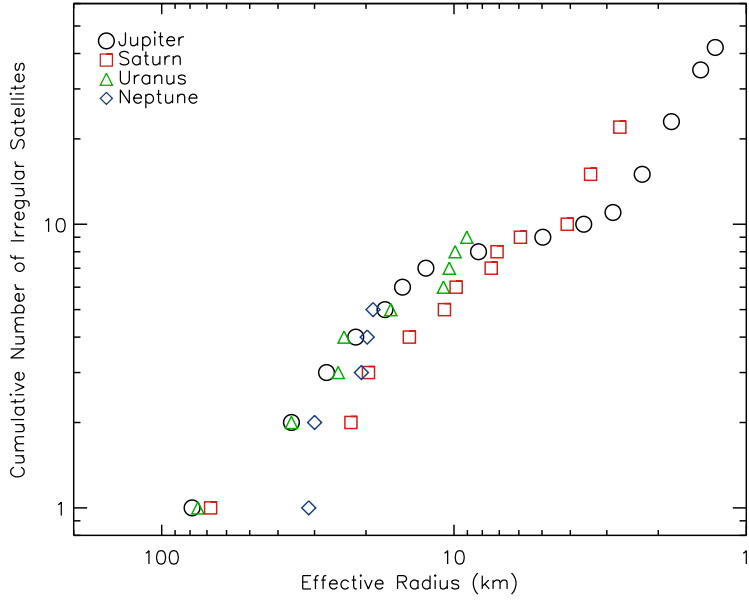}
\vspace{1cm}
\caption{The cumulative radius function for the irregular satellites
with $r < 100$ km of Jupiter, Saturn, Uranus and Neptune.  This figure
directly compares the sizes of the satellites of all the giant planets
assuming all satellite populations have albedos of about 0.04.
Jupiter, Saturn and Uranus all have shallow irregular satellite size
distributions of $q \sim 2$ for satellites with $100 > r > 10$ km.
Neptune's limited number of known small outer irregular satellites
with $100 > r > 10$ km show a steeper size distribution of $q \sim 4$,
but if Nereid and/or Triton are included we find a much shallower size
distribution of $q\sim 1.5$.  Both Jupiter and Saturn appear to show a
steeper size distribution for irregular satellites with $r < 5$ km
which may be a sign of collisional processing.  To date neither
Uranus' or Neptune's Hill spheres have been surveyed to these smaller
sizes.  Further discoveries of irregular satellites around Neptune are
needed to obtain a reliable size distribution.}
\label{fig:cumneptsize} 
\end{figure}

We model the irregular satellite size distribution through a
differential power-law radius distribution of the form $n(r)dr=\Gamma
r^{-q}dr$, where $\Gamma$ and $q$ are constants, $r$ is the radius of
the satellite, and $n(r)dr$ is the number of satellites with radii in
the range $r$ to $r+dr$.  The slope of the CLF ($\alpha$) and exponent
of the size distribution ($q$) are simply related by $q = 5 \alpha +
1$ when assuming similar heliocentric distance and albedos for all
satellites.  We show the size distribution of outer irregular
satellites with $r< 100$ km in Figure \ref{fig:cumneptsize}.  Using
$\alpha = 0.6$ for Neptune's outer satellites with $r< 100$ km we find
$q \sim 4$ but if we include Nereid and/or Triton in which $\alpha
\sim 0.06$ we find $q \sim 1.3$.  Because of the large sensitivity on
Nereid, these results are still consistent with the shallow $q \sim 2$
found for the irregular satellites with $10 < r < 100$ km of Jupiter,
Saturn and Uranus (Sheppard \& Jewitt 2003; Kavelaars et al. 2004;
Sheppard et al. 2005).  Jupiter's smallest satellites ($r < 5$ km)
follow a steeper power law of $q \sim 3.5$ while Saturn's small
irregulars also show a steepening in the size distribution for $r < 5$
km.  Uranus' known irregulars do not yet extend down to these small
sizes (see Table 2).  These ``bumps'' in the size distribution are
probably caused by the collisional evolution of the irregular
satellites and may be similar to what has been observed in the main
belt of asteroids (Davis et al. 2002; Bottke et al. 2005).  The large
($r>50$ km) Kuiper Belt objects and Centaurs both have similar size
distributions of $q \sim 4$ (Trujillo et al. 2001; Sheppard et
al. 2000) while the smaller Kuiper Belt objects may have a shallower
slope (Bernstein et al. 2004).  Smaller Neptune satellites probably
await discovery and will allow us to determine if the steep size
distribution power law continues to these smaller objects.  If the
slope is significantly different than that found for the other giant
planets it may be a disruption signature from the capture of Triton
and the scattering of Nereid from the regular satellite population.

\section{Summary}

1) We have conducted an ultra deep survey of 1.75 deg$^2$ around
Neptune reaching $50 \%$ detection efficiency at a red limiting
magnitude of 25.8 mags.  This corresponds to objects with $r >$ 17 km
(for an assumed albedo of 0.04).

2) We discovered one new satellite, Psamathe (S/2003 N1), and detected
four of five previously known small irregular satellites in our
survey.  Psamathe is about 20 km in radius (assuming an albedo of
0.04) and has a distant, eccentric retrograde orbit similar to those
of other irregular satellites thought to have been acquired by
capture.

3) Neptune has a distant irregular satellite population with sizes and
 orbital  properties  like those  of  the  irregular satellites  found
 around Jupiter, Saturn and Uranus.

4) The size distribution of Neptune's irregular satellites is poorly
determined by the existing data.  Larger samples of the small outer
irregular satellites of Neptune are needed to determine the size
distribution with more confidence.  Shallow power law size
distributions have been found for the irregular satellites with $100 >
r > 10$ km around Jupiter, Saturn and Uranus ($q \sim 2$) while
steeper power laws ($q \sim 3.5$) appear for satellites with $r<5$ km
which may be a sign of collisional evolution.

\section*{Acknowledgments}

We thank Brian Marsden and Bob Jacobson for orbit determinations of
the satellites.  This work was partially supported by a NASA grant to
DJ.  Additional support for this work was provided by NASA through
Hubble Fellowship grant \# HF-01178.01-A awarded by the Space
Telescope Science Institute, which is operated by the Association of
Universities for Research in Astronomy, Inc., for NASA, under contract
NAS 5-26555.

\newpage



\begin{center}
\begin{deluxetable}{lccccc}
\tablenum{1}
\tablewidth{6 in}
\tablecaption{Geometrical Circumstances of Neptune}
\tablecolumns{6}
\tablehead{
\colhead{UT Date} & \colhead{R}  & \colhead{$\Delta$} & \colhead{$\alpha$} & \colhead{RA} & \colhead{Dec} \\ \colhead{} &\colhead{(AU)} &\colhead{(AU)} &\colhead{(deg)} & \colhead{($\arcsec /hr$)} & \colhead{($\arcsec / hr$)} }  
\startdata
2003 Aug 29 & 30.078 & 29.156 & 0.79 & -3.5 & -1.0 \nl
2003 Aug 30 & 30.078 & 29.163 & 0.82 & -3.5 & -1.0 \nl
\enddata
\end{deluxetable}
\end{center}


\newpage



\begin{center}
\begin{deluxetable}{cccccc}
\tablenum{2}
\tablewidth{6 in}
\tablecaption{Outer Irregular Satellites of the Planets}
\tablecolumns{6}
\tablehead{
\colhead{Planet} & \colhead{Irr\tablenotemark{a}}  & \colhead{R-Mag\tablenotemark{b}} & \colhead{$r_{min}$\tablenotemark{c}} & \colhead{Hill\tablenotemark{d}} & \colhead{Hill} \\ \colhead{} &\colhead{} &\colhead{limit} &\colhead{limit} & \colhead{Radii} & \colhead{Radii} \\ \colhead{} & \colhead{(\#)} & \colhead{(mag)} & \colhead{(km)} & \colhead{(deg)} & \colhead{($10^{7}$km)} }  
\startdata
Mars\tablenotemark{e}   & 0    & 23.5  & 0.1  & 0.7  & 0.1 \nl
Jupiter & 55   & 23.5  & 1    & 4.7  & 5.1  \nl
Saturn  & 26   & 24.5  & 3    & 3.0  & 6.9 \nl
Uranus  & 9    & 26    & 7    & 1.4  & 7.3  \nl
Neptune\tablenotemark{f} & 5(7) & 25.5  & 20   & 1.5  & 11.6  \nl
\enddata
\tablenotetext{a}{Number of known outer irregular satellites as of December 1, 2005.}
\tablenotetext{b}{Approximate limitting magnitude in the R-band of completeness for respective planet's outer satellites.}
\tablenotetext{c}{Approximate limiting radii of satellite searches to date.}
\tablenotetext{d}{The apparent angular Hill Sphere radius of the planet at opposition.}
\tablenotetext{e}{Mars' two inner satellites may have been captured in a similar way as the outer irregular satellites of the giant planets.}
\tablenotetext{f}{Neptune only has 5 ``normal'' irregular satellites if the ``unusual'' Triton and Nereid are not included.}
\end{deluxetable}
\end{center}


\newpage



\begin{center}
\begin{deluxetable}{lccccccccc}
\tablenum{3}
\tablewidth{6.5 in}
\tablecaption{Physical and Orbital Properties of Neptune's Irregular Satellites\tablenotemark{*}}
\tablecolumns{10}
\tablehead{
\colhead{Name} & \colhead{a\tablenotemark{a}} & \colhead{i\tablenotemark{b}} & \colhead{e\tablenotemark{c}} & \colhead{Peri\tablenotemark{d}} & \colhead{Node\tablenotemark{e}} & \colhead{M\tablenotemark{f}} & \colhead{Period\tablenotemark{g}} & \colhead{mag.\tablenotemark{h}} & \colhead{r\tablenotemark{i}} \\ \colhead{} &\colhead{($10^{3}$km)} &\colhead{(deg)} &\colhead{} & \colhead{(deg)} & \colhead{(deg)} & \colhead{(deg)} & \colhead{(days)} & \colhead{($m_{R}$)} & \colhead{(km)} }  
\startdata
I  Triton\tablenotemark{j} &    355  &  157  &  0.00  & 344.0  &  172.4  &  264.8  &   5.88   &  13.0  &  1353  \nl
II Nereid\tablenotemark{k} &   5513  &  7.2  &  0.75  & 280.8  &  334.8  &  359.3  &   360.1  &  19.2  &  170   \nl
           S/2002 N1       &  15728  &  134  &  0.57  & 159.7  &  203.0  &  96.4   &   1879.7 &  24.5  &  31    \nl
           S/2002 N2       &  22422  &  48   &  0.29  &  79.3  &  55.5   &  207.1  &   2914.1 &  25.5  &  22    \nl
           S/2002 N3       &  23571  &  35   &  0.42  & 142.4  &  60.7   &  328.6  &   3167.9 &  25.5  &  21    \nl
           S/2003 N1       &  46695  &  137  &  0.45  & 145.9  &  301.0  &  206.2  &   9115.9 &  25.5  &  20    \nl
           S/2002 N4       &  48387  &  133  &  0.49  &  89.3  &   50.0  &  269.8  &   9374.0 &  24.6  &  30    \nl
\enddata
\tablenotetext{*}{Orbital data are from Robert Jacobson at JPL ($http://ssd.jpl.nasa.gov/?sat\_elem$), fits are over a 1000 year time span. Epochs are June 10, 2003 TT for the Neptune satellites discovered in 2002 and 2003 and Aug 25, 1989 TT for Triton and Nereid.}
\tablenotetext{a}{Mean semi-major axis with respect to Neptune.}
\tablenotetext{b}{Mean inclination of orbit with respect to the ecliptic.}
\tablenotetext{c}{Mean eccentricity.}
\tablenotetext{d}{The argument of Pariaphis.}
\tablenotetext{e}{The longitude of the ascending node.}
\tablenotetext{f}{The mean anomaly.}
\tablenotetext{g}{Orbital period of satellite around Neptune.}
\tablenotetext{h}{Apparent red (0.65 $\micron$ wavelength) magnitude. Uncertainties are around 0.2 mags.}
\tablenotetext{i}{Radius of satellite assuming a geometric albedo of 0.04.}
\tablenotetext{j}{Triton is an ``unusual'' retrograde satellite and not classified as an irregular satellite under the definition of Burns (1986).  Triton likely has had significant modification of its orbit from tidal interactions with Neptune (Goldreich et al. 1989; Chyba et al. 1989).  Triton's mean inclination as shown here is with respect to Neptune's equator.}
\tablenotetext{k}{Nereid is an ``unusual'' irregular satellite because of its relatively low inclination, small semi-major axis and large eccentricity.  It may be a perturbed regular satellite.}
\end{deluxetable}
\end{center}


\newpage

\clearpage

\clearpage

\clearpage

\clearpage

\clearpage

\clearpage

\clearpage

\clearpage

\end{document}